\documentclass[preprint]{aastex}
\usepackage{graphicx}
\usepackage{array}
\usepackage{longtable}
\usepackage{natbib}
\usepackage{amssymb}
\begin{document}

\title{Super-Chandrasekhar SNe Ia Strongly Prefer Metal-Poor Environments}

\author{Rubab~Khan\altaffilmark{1},
K.~Z.~Stanek\altaffilmark{1,2},
R.~Stoll\altaffilmark{1},
J.~L.~Prieto\altaffilmark{3,4}
}

\altaffiltext{1}{Dept.\ of Astronomy, The Ohio State University, 140
W.\ 18th Ave., Columbus, OH 43210; khan, kstanek, stoll@astronomy.ohio-state.edu}

\altaffiltext{2}{Center for Cosmology and AstroParticle Physics, 
The Ohio State University, 191 W.\ Woodruff Ave., Columbus, OH 43210}

\altaffiltext{3}{Carnegie Observatories, 813 Santa Barbara Street, 
Pasadena, CA 91101; jose@obs.carnegiescience.edu}

\altaffiltext{4}{Hubble, Carnegie-Princeton Fellow.}

\shorttitle{The Brightest Mid-IR Source in M33}

\shortauthors{Khan et al.~2010}

\begin{abstract}
\label{sec:abstract}

We discuss the emerging trend that super-Chandrasekhar Type~Ia Supernovae 
(SCSNe) with progenitor mass estimates significantly exceeding $\sim1.4~M_\odot$ 
tend to explode in metal-poor environments. While \cite{ref:Taubenberger_2011} 
noted that some of the SCSNe host galaxies are relatively metal-poor, we focus 
quantitatively on their locations in the hosts to point out that in three out of 
four cases, the SCSNe explosions occurred in the outer edge of the disks 
of their hosts. It is therefore very likely that their 
progenitors had far lower metallicities than those implied by the metallicity 
of their hosts' central regions. In two cases (SN~2003fg and SN~2009dc) the 
explosion sites were outside $\sim99\%$ of the host's light, and in one case 
(SN~2006gz) the host's radial metallicity slope indicates that the explosion 
site is in a metal-poor region. The fourth case (SN~2007if) has the lowest 
spectroscopically measured SN~Ia host metallicity~\citep{ref:Childress_2011}. 
It may be possible to explain each 
of these unusually bright events through some progenitor scenario specific to 
that case, but a much simpler and straightforward conclusion would be to ascribe 
the controlling factor to the only physical aspect they have in common --- 
metal poor environments. 

\end{abstract} 
\keywords{supernovae: general, supernovae: individual: SNe 2003fg 2006gz 2007if 2009dc, white dwarfs}
\maketitle

\section{Introduction}
\label{sec:introduction}

Type Ia Supernovae (SNe~Ia) are important both as astrophysical probes of 
cosmology \citep[e.g.,][and references therein]{ref:Kessler_2009} and for their 
role in the chemical evolution of galaxies~\citep[e.g.,][]{ref:Kobayashi_2009}. 
At the same time, the nature of SNe~Ia progenitors and their explosion 
mechanism are still uncertain. Two possible explosion scenarios have been proposed. 
In the single degenerate (SD) scenario~\citep[e.g.,][]{ref:Whelan_1973,ref:Piro_2008,ref:Meng_2011a}, 
an accreting carbon/oxygen white dwarf (CO WD) undergoes thermonuclear 
explosion when its mass approaches the Chandrasekhar limit of 
$\sim1.4~M_\odot$~\citep{ref:Chandrasekhar_1931}. In the double degenerate (DD) 
scenario, two CO WDs merge, possibly with the help of a third star in the system 
accelerating the merger through the Kozai mechanism~\citep{ref:Thompson_2010}, 
to produce a progenitor that potentially is 
$>1.4~M_\odot$~\citep[e.g.,][]{ref:Iben_1984,ref:Webbink_1984,ref:Pakmor_2010}. 
The decay of $^{\rm 56}$Ni produced by SN~Ia explosion and the subsequent decay 
product $^{\rm 56}$Co power the observed luminosity~\citep{ref:Colgate_1969}.

Over the last decade, at least four SNe~Ia have been discovered that were too 
luminous to have resulted from thermonuclear explosions of Chandrasekhar-mass 
WDs. It has been proposed, first by \citet{ref:Howell_2006} for the case 
of SN~2003fg, that these four SNe~Ia exploded through DD mergers. While it 
requires $\sim0.6~M_\odot$ of $^{\rm 56}$Ni to power a normal SN~Ia lightcurve 
with a progenitor mass $\sim1.4~M_\odot$~\citep[e.g.,][]{ref:Nomoto_1984,ref:Branch_1992}, 
the inferred $^{\rm 56}$Ni masses of the four super-Chandrasekhar SNe~Ia range 
from $\sim1.2~M_\odot$ to $\sim1.7~M_\odot$ and imply total progenitor masses 
from $\sim2.0~M_\odot$ to $\sim2.4~M_\odot$ (see Table~1). 

\cite{ref:Taubenberger_2011} noted that some of the SCSNe host galaxies are 
relatively metal-poor. In our discussion of SN~2009nr~\citep{ref:Khan_2011}, 
we pointed out that for three of the four confirmed SCSNe events, the SNe 
were located far from the centers of their host galaxies, and probably were 
in significantly lower metallicity environments than implied by the metallicity 
of their hosts' central regions. The fourth case, SN~2007if, is located in a 
low-luminosity dwarf galaxy that has the lowest spectroscopically measured 
metallicity among SN~Ia hosts (12+log(O/H)=$8.01\pm0.09$ or $Z \simeq 0.15 Z_\odot$, 
\citealt{ref:Childress_2011}, using the $R_{23}$ method of \citealt{ref:Kobulnicky_2004}). 
This motivated us to further examine the currently available evidence that 
SCSNe show a strong preference for metal-poor environments, 
quantitatively focusing on their locations in their hosts. Section~\ref{sec:data} 
describes the SNe host galaxy data and Section~\ref{sec:analysis} presents our 
analysis of the host properties. In Section~\ref{sec:discussion}, we consider 
the implications of our results and their consequences for SNe~Ia progenitor 
theories.

\section{Data}
\label{sec:data}

\begin{figure*}[!t]
\begin{center}
{\includegraphics[width=135mm]{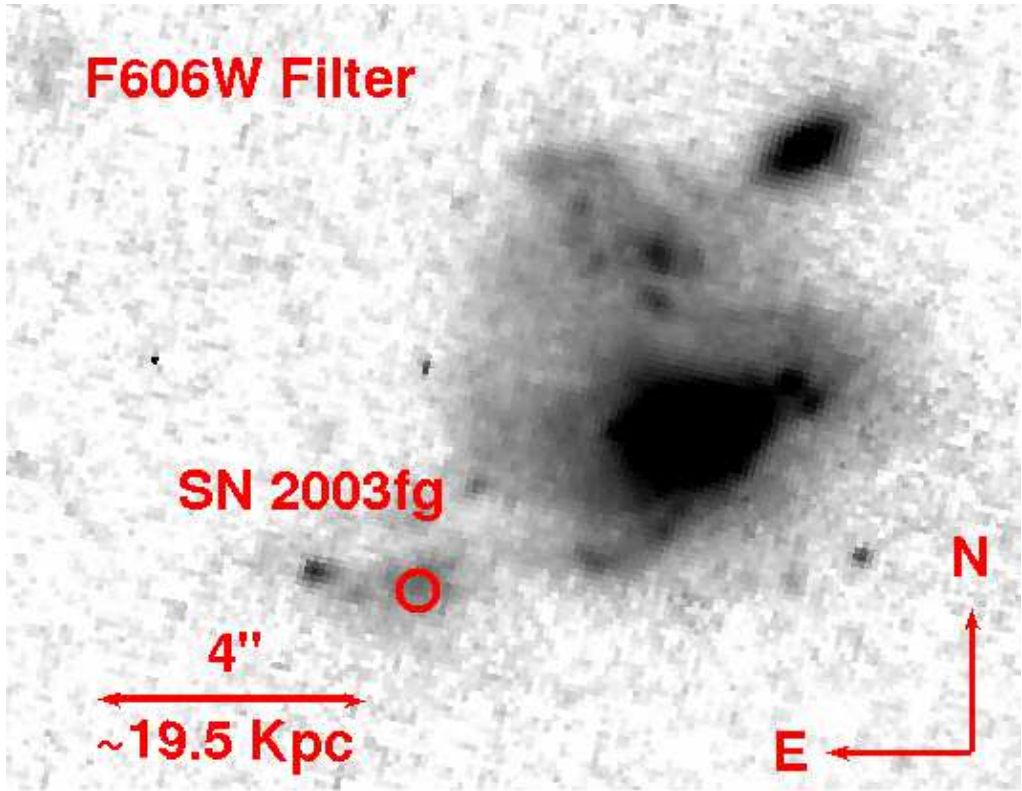}}
\end{center}
\caption{A HST WFPC2 view of the SN~2003fg explosion site in the F606W filter. 
The circle marks the location of the SN. The center of the presumed host 
\citep{ref:Howell_2006} is inside the circle, although it could plausibly be a 
tidal feature of the large morphologically disturbed galaxy at the same redshift.}
\label{fig:sn03fg}
\end{figure*}

\begin{figure*}[!t]
\begin{center}
{\includegraphics[width=135mm]{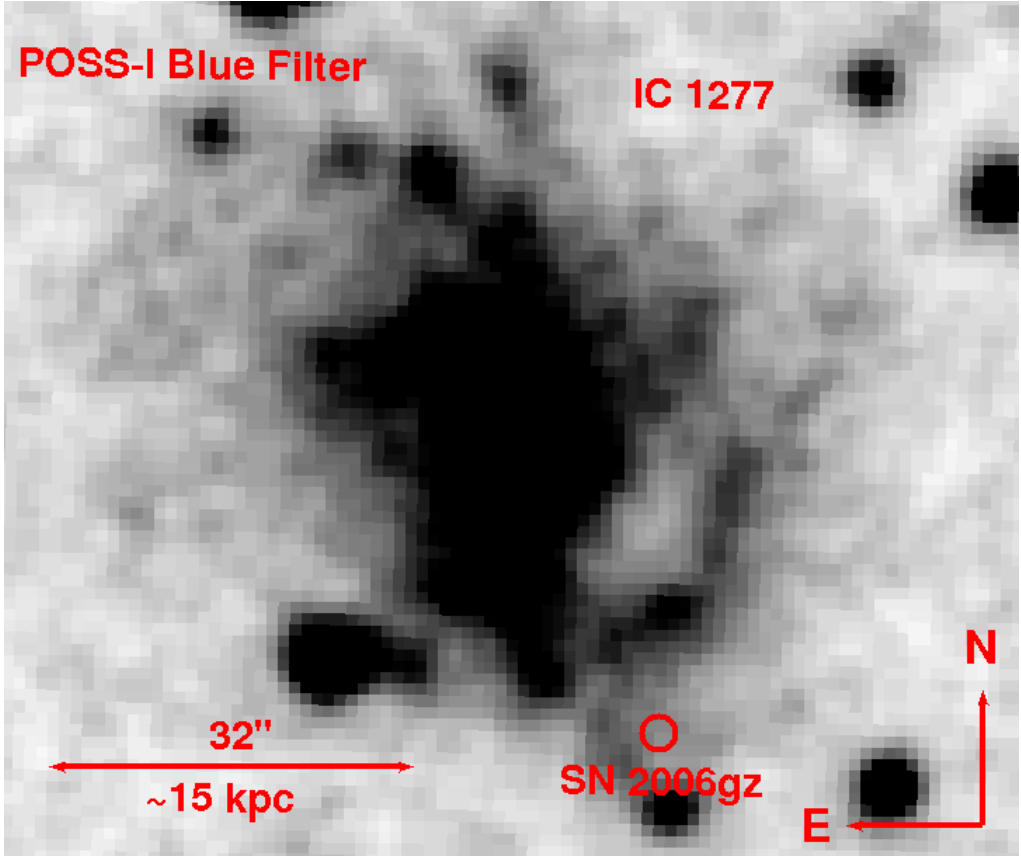}}
\end{center}
\caption{A Palomar Schmidt telescope image of the SN~2006gz host galaxy IC~1277 
taken through the POSS-I blue filter. The location of the SN is marked with a circle.}
\label{fig:sn06gz}
\end{figure*}

\begin{figure*}[!t]
\begin{center}
{\includegraphics[width=135mm]{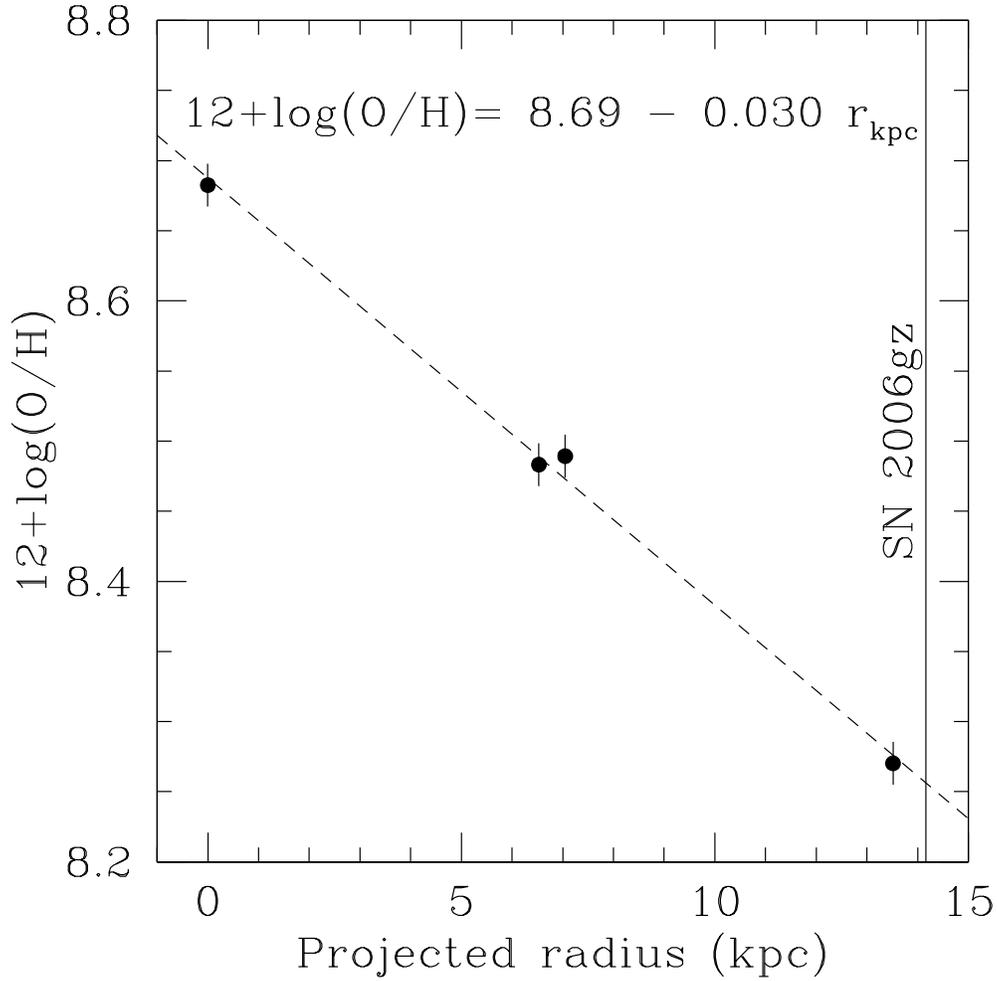}}
\end{center}
\caption{The radial oxygen abundance profile of the SN~2006gz host galaxy 
IC~1277 along a slit going through the SN and the center of the host (based on 
four off-center \ion{H}{2} regions). The dashed line shows a linear fit to 
the metallicity measurementsand has a gradient of $\sim0.03$~dex~kpc$^{-1}$. 
The solid line marks the position of the SN and intersects the dashed line at 
12+log(O/H)$\simeq8.26$.}
\label{fig:slope}
\end{figure*}

\begin{figure*}[!t]
\begin{center}
{\includegraphics[width=135mm]{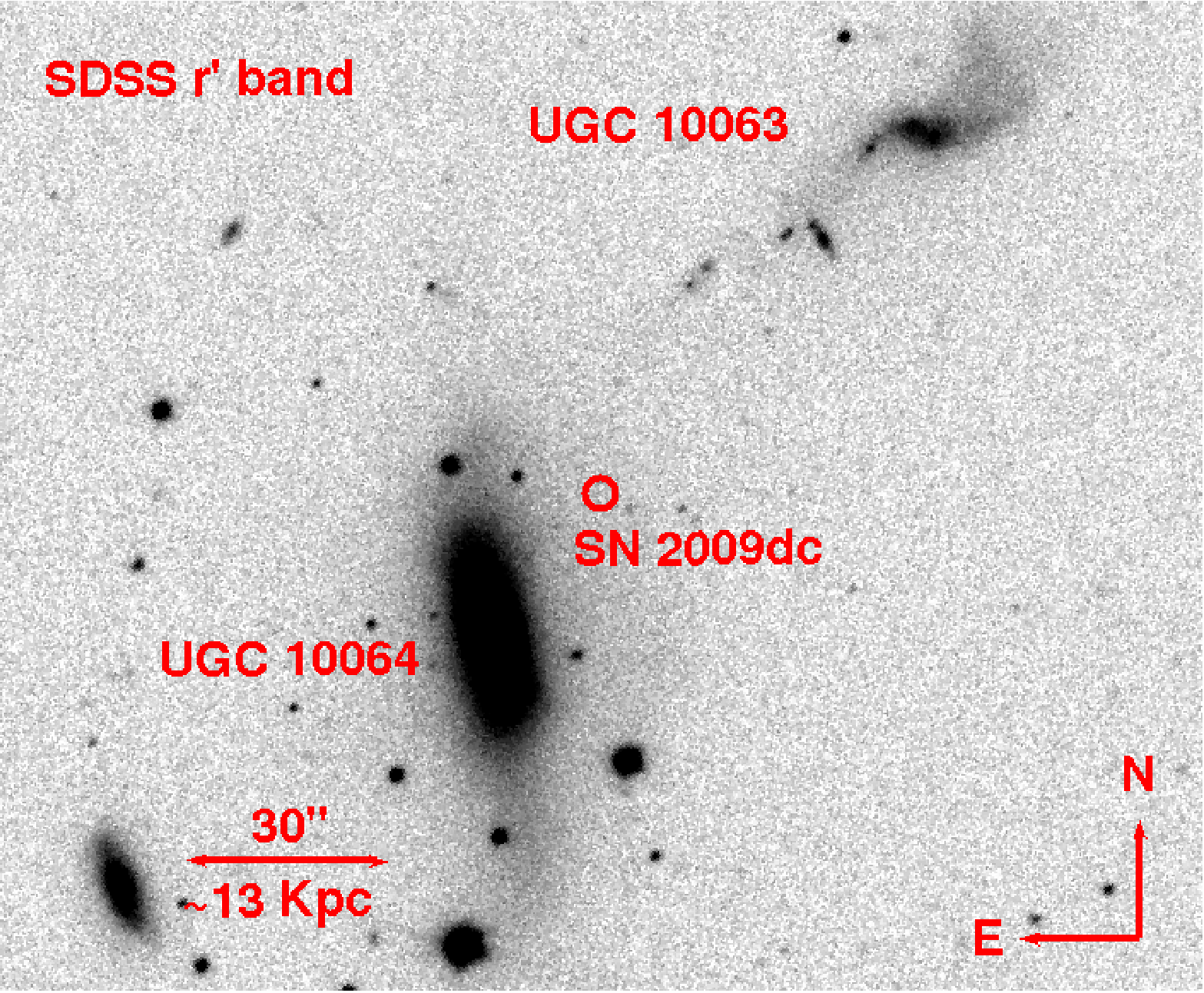}}
\end{center}
\caption{A SDSS $r'$-band image of the SN~2009dc host galaxy UGC~10064. The 
other galaxy at the top right, UGC~10063, is at the same redshift, and the the 
two galaxies may be linked by a tidal bridge~\citep{ref:Taubenberger_2011}. 
The location of the SN is marked by a circle.}
\label{fig:sn09dc}
\end{figure*}

We gathered publicly available images of the four SCSNe explosion sites in 
order to quantitatively characterize their host properties. For SN~2003fg, 
five Hubble Space Telescope (HST) WFPC2 images taken through the F606W filter 
\citep[HST Prog. ID 8698, PI: J.~Mould;][]{ref:Vogt_2005} were obtained from 
the HST archive\footnote{\scriptsize{\tt http://archive.stsci.edu/hst/}}, and 
they were combined to produce Figure \ref{fig:sn03fg}. For SN~2006gz, a Palomar 
Schmidt telescope image of the host galaxy IC~1277 taken through the POSS-I 
blue filter (Figure \ref{fig:sn06gz}) was obtained from the STScI Digitized Sky 
Survey archive\footnote{\scriptsize{\tt http://archive.stsci.edu/cgi-bin/dss\_form}}. 
For  SN~2009dc, a Sloan Digital Sky Survey (SDSS) \textit{r\'} band image of 
the host galaxy UGC~10064 (Figure \ref{fig:sn09dc}) was obtained from the SDSS 
Data Release 7 archive\footnote{\scriptsize{\tt http://www.sdss.org/dr7/}}. 
For the SN~2007if host properties, we use the results presented by 
\citet{ref:Childress_2011}.

Spectra of the SN~2006gz host galaxy IC~1277 and the large galaxy near 
the presumed dwarf host of SN~2003fg were obtained on 2011 April 6 with 
the Ohio State Multi-Object Spectrograph 
\citep[OSMOS;][]{ref:Stoll_2010,ref:Martini_2011} on the 2.4-m Hiltner telescope. 
The OSMOS slit was oriented N/S and was centered on the galaxies. The raw data were 
processed using standard techniques in IRAF\footnote{IRAF is distributed by the 
National Optical Astronomy Observatory, which is operated by the Association of 
Universities for Research in Astronomy (AURA) under cooperative agreement with 
the National Science Foundation.}, including cosmic ray rejection using 
L.~A.~Cosmic (van Dokkum 2001). The slit location (the red-sensitive inner 
slit\footnote{\scriptsize{\tt http://www.astronomy.ohio-state.edu/~martini/osmos/}}) 
we used provides a wavelength coverage of 3960-6880 \AA. The spectra were 
calibrated in wavelength using Ar arclamp spectra and in flux using coincident 
observations of a spectrophotometric standard from Oke (1990). 

For IC~1277, we extracted spectra of four HII regions along the slit using apertures 
of 14 pixels (3.8 arcsec). We determined their oxygen abundances using the N2 diagnostic of 
\cite{ref:Pettini_2004}, which depends solely on 
[\ion{N}{2}]$\lambda 6584$/H$\alpha \lambda 6563$. Due to the proximity of the 
lines, this ratio is very insensitive to reddening. For the galaxy near SN~2003fg, 
given the wavelength range at this redshift, only R23 methods were available to 
determine metallicity from the spectrum taken at its center, and estimates from the 
R23 methods were converted to the N2 scale that we used for IC~1277 using the empirical 
conversions of~\cite{ref:Kewley_2008}. The R23 methods are very 
sensitive to reddening correction, and we corrected only for Galactic reddening using 
E(B-V)=0.013~mag~\citep{ref:Schlegel_1998}, using the CCM function, and assuming R=3.1.

\section{Analysis}
\label{sec:analysis}

We used GALFIT~\citep{ref:Peng_2002} to model the host galaxies. We fit simple 
exponential disk profiles to the large morphologically disturbed galaxy near 
the presumed dwarf host of SN~2003fg (see Figure~\ref{fig:sn03fg}) and to the 
host of SN~2006gz (IC~1277). For SN~2009dc, we fit a two-component (exponential 
plus sersic) disk/bulge decomposition profile to the S0 host UGC~10064. We also 
used the IRAF$^4$ Ellipse tool to analyze the galaxy profiles produced by 
GALFIT to approximately determine what fraction of total light is located 
interior to the location of the SNe in the hosts. Finally, we determined the 
radial metallicity slope of the host galaxy of SN~2006gz (IC~1277) from the 
spectroscopic data described in Section~\ref{sec:data}. A summary of the 
properties of the SCSNe events and their host galaxies is presented in Table~1. 
Throughout this section, we assume $H_0=72$, $\Omega_M=0.27$, and a flat 
universe when translating angular size or separation (arcsecs) to physical 
distance (kpc).

SN~2003fg is located $4\farcs04$E and $2\farcs53$S from the center of the 
morphologically disturbed neighbouring galaxy (projected distance 
$\sim23.5$~kpc; see Figure~\ref{fig:sn03fg}), which is at the same redshift 
as the presumed dwarf host, although the dwarf galaxy could also be a tidal 
feature of the much larger neighbour~\citep{ref:Howell_2006}. The explosion 
site is $\sim0.9$~kpc away from the center of the dwarf host, and thus under 
any assumption the SN would be located in a metal-poor environment if it is 
indeed associated with this dwarf galaxy. Alternatively, the parameters
derived by GALFIT indicate that the larger neighbour has an exponential 
disk scale length of $\sim0\farcs64$ or $\sim3.2$~kpc. If this galaxy is the 
actual host, then the SN exploded $\gtrsim7$ disk scale lengths away from the 
center of its host. The GALFIT profile of this galaxy indicates that $\sim99\%$ 
of its light is located interior to the location of the SN. An exponential disk 
profile is likely not the optimal function to model disturbed galaxies such as 
this, however there does not appear to be any significant flux excess towards the location 
of the SN except for the dwarf galaxy-like feature, and our estimate for 
the fraction of light within the SN location appear robust. Metallicity estimates 
of the center of this galaxy from the spectroscopic data presented in 
Section~\ref{sec:data} using various R23 methods convert to a consistent value 
of 12+log(O/H)$\simeq8.57$ ($Z \simeq 0.5 Z_\odot$) on the N2 diagnostic scale 
of \cite{ref:Pettini_2004}. 
Although the metallicity gradient of morphologically disturbed galaxies is uncertain, 
our measurement shows that SN~2003fg is likely associated with a low-metallicity 
environment even if it originated from material ejected from the large galaxy with 
metallicity as high as that of its central region. 

SN~2006gz is located $12''$W and $28''$S from the center of the Scd host 
IC~1277 at a projected distance of 14.4~kpc~\citep{ref:Hicken_2007}. The 
parameters derived by GALFIT show that IC~1277 has an exponential disk scale 
length of $\sim14\farcs4$ or 6.7~kpc. This means that the SN exploded 
$\sim2.1$ disk scale lengths away from the center of the host at the 
outer edge of the disk of the galaxy 
(Figure~\ref{fig:sn06gz}). The SN is located outside of $\sim75\%$ of 
the galaxy's light. Analyzing the spectroscopic data presented in 
Section~\ref{sec:data}, we determined the radial oxygen abundance profile of 
the host. A linear fit to the oxygen abundance measurements has a gradient of 
$\sim0.03$~dex~kpc$^{-1}$. In Figure~\ref{fig:slope}, the dashed line shows the 
linear fit to the metallicity measurements and the solid line marks the 
position of the SN. It is apparent that the SN location is metal-poor 
(12+log(O/H)$\simeq8.26$ or $Z \simeq 0.25 Z_\odot$).

\begin{table}[p]
\begin{center}
\label{table:scsn}
\begin{tabular}{lcccc}
\hline 
\hline
\\
\multicolumn{1}{l}{Property} &
\multicolumn{1}{c}{SN~2003fg} &
\multicolumn{1}{c}{SN~2006gz} &
\multicolumn{1}{c}{SN~2007if} &
\multicolumn{1}{c}{SN~2009dc}
\\
\hline
\hline
\\
RA & 14$^h$16$^m$18\fs8 & 18$^h$10$^m$26\fs3 & 1$^h$10$^m$51\fs4 & 15$^h$51$^m$12\fs1 \\
Dec  & +52\arcdeg14\arcmin55 & +30\arcdeg59\arcmin44 & +15\arcdeg27\arcmin40 & +25\arcdeg42\arcmin28\\
Redshift & $0.2440\pm0.0003$ & $0.0234\pm0.0004$ & $0.07416\pm0.00082$ & $0.022\pm0.001$\\
$M_V$ (SN)  & $-20.0$ & $-19.2$ & $-20.4$ & $-19.8$\\
$\Delta m_{15}$ (B)  & 0.94 & $0.69\pm0.04$ & $0.71\pm0.06$ & $0.65\pm0.03$\\
$^{\rm 56}$Ni Mass & $1.3~M_\odot$ & $1.2~M_\odot$ & $1.6~M_\odot$ & $1.7~M_\odot$\\
M$_{tot}$ & $2.1~M_\odot$ & --- & $2.4~M_\odot$ & $\gtrsim 2~M_\odot$\\
\\
\hline
\\
Host Name  & Unnamed & IC~1277 & Unnamed & UGC~10064\\
Host Magnitude  & $M_{F606W}=-21.3$ & $M_B=-20.3$ & $M_g=-14.1$ & $M_z=-21.9$\\
Axis raio (b/a) & $\sim 0.6$ & $\sim 0.7$ & --- & $\lesssim 0.1$\\
D$_1$ & $4\farcs8$ & $30\farcs5$ & --- & $27''$\\
D$_2$ & 23.5 kpc & 14.4 kpc & --- & 12 kpc \\
L($<$r)  & $\sim99\%$ & $\sim75\%$ & --- & $\gtrsim99\%$ 
\\
\hline
\hline
\end{tabular}
\end{center}
\caption{Summary of the SCSNe events and their host properties. The SNe 
properties (coordinates, redshift, peak magnitude, light-curve decline 
rate, derived $^{\rm 56}$Ni mass and total progenitor mass) are from 
\cite{ref:Howell_2006}, \cite{ref:Hicken_2007}, \cite{ref:Yamanaka_2009}, 
\cite{ref:Scalzo_2010}, and \cite{ref:Silverman_2011}. The host magnitudes are 
from the RC3 catalog~\citep{ref:RC3_1991} for SN~2006gz, \cite{ref:Scalzo_2010} 
for SN~2007if, and the SDSS DR6 catalog~\citep{ref:Adelman_2008} for SN~2009dc. 
As the SN~2003fg host properties, we present those of the morphologically 
disturbed large neighbouring galaxy at the same redshift as the presumed dwarf 
host, and the absolute magnitude was determined from its GALFIT profile. The 
exponential disk scale axis ratios are from our GALFIT results, while D$_1$ 
(arcseconds) and D$_2$ (kiloparsecs) are distances of the SNe explosion sites 
from the centers of their hosts. L($<$r) is the approximate percentage of light 
located interior to the location of the SN in its host. No total progenitor 
mass estimate for SN~2003fg was made by \cite{ref:Hicken_2007}. SN~2007if 
exploded at the center of its faint low metallicity dwarf host.}
\end{table}

SN~2009dc is located $15\farcs8$W and $20\farcs8$N of the S0 host galaxy 
UGC~10064 (Figure~\ref{fig:sn09dc}) at a projected distance of 12~kpc 
\citep{ref:Silverman_2011}. The parameters derived from GALFIT show that 
the host has an exponential disk scale length of $\sim13\farcs8$ 
($\sim6$~kpc) and its bulge (Sersic index $\sim3.8$) has an effective radius 
of $\sim8''$ ($\sim3.5$~kpc). This indicates that the SN exploded $\sim2$ 
exponential disk scale lengths away from center of UGC~10064. The host is very 
highly inclined with GALFIT estimates for the axis ratio of $\lesssim0.1$ for 
both the disk and the bulge components. If the SN exploded in an extended disk, 
the actual distance from the center of the host could be more than ten times 
larger, by simple geometric considerations. Although the SN is located outside 
of $\gtrsim99\%$ of the galaxy's light, the SDSS $g'$ and $r'$ band images show 
low surface brightness emission at the position of the SN that is fairly 
asymmetric around the galaxy, with more light in the side where the SN 
is (NW) compared to the opposite side (SE). This appears consistent with a 
fairly inclined disk, perhaps an extended disk. It has also been proposed by 
\cite{ref:Taubenberger_2011} that SN~2009dc might be associated with a faint 
and narrow tidal stream connecting the host with the $\sim10$ times less massive 
neighboring galaxy UGC~10063 and the progenitor system may have formed a few hundred Myr ago 
during an interaction of these two galaxies. 
It is impossible to measure the gas phase oxygen abundance or the metallicity 
gradient of this gas-poor galaxy directly using gas-phase abundances as we did for the 
host of SN~2003fg, but the central region of UGC~10064 appears to have a relatively 
undisturbed S0 profile despite being in an interacting system, and it is reasonable 
to expect a metallicity gradient fairly similar to that of a relatively undisturbed 
spiral galaxy. Since during a galaxy interaction it is easier to disturb mass that 
is not deep in the potential well of the significantly more massive galaxy, even 
if SN~2009dc originated from mass ejected from the host, it is likely associated with 
low-metallicity material stripped from the outer regions.

\section{Discussion}
\label{sec:discussion}

Several papers published recently have discussed low metallicity as an 
important ingredient for the progenitors of possible SCSNe explosions. 
\cite{ref:Silverman_2011} noted that while the host galaxy types and 
distances of the SCSNe explosion sites from the centers of their hosts 
vary, a number of the SCSNe exploded far from the centers of the host 
galaxies. \cite{ref:Childress_2011} pointed out many unusual 
SNe~Ia, including the SCSNe, have been discovered in low luminosity hosts. 
\cite{ref:Taubenberger_2011} concluded that 
SCSNe show a tendency to explode in low-mass galaxies, and low-metallicity 
progenitors may be an important prerequisite for producing superluminous 
SNe~Ia. \cite{ref:Khan_2011} observed that three of the four confirmed SCSNe 
events were located far from the centers of their host galaxies, probably in 
significantly lower metallicity environments than implied by the metallicity of 
their hosts' central regions. In the current work, we have quantitatively focused 
on their locations in the hosts to show that all of the 
SCSNe progenitors appear to be associated with regions of their hosts likely to 
be metal-poor. It is also important to note that the projected distances are only lower 
limits on the physical separations between the SN explosion sites and the centers of their 
hosts, and the actual distances will be larger and hence likely will have still lower metallicity. 
One could argue for a unique explanation for each of these four events, specially 
given that in two cases (SN~2003fg and SN~2009dc) the SNe are associated with 
interacting systems, but Occam's Razor suggests that it is simply due to a 
characteristic they all share, such as low metallicity.

Although a detailed study of the stellar populations at the outer edges 
of the SCSNe host galaxies is not possible, studies of Local Group 
galaxies demonstrate that the outer regions of galaxies primarily contain 
metal-poor stars. The halo of the Milky Way is dominated by metal-poor 
stellar populations~\citep{ref:Brown_2010}. In case of the M31 spheroid, 
the metallicity is known to decrease monotonically with increasing radius 
--- it has mean [Fe/H] values of $-0.65$ at 11~kpc, $-0.87$ at 21~kpc, and 
$-0.98$ at 35 kpc from the nucleus~\citep{ref:Brown_2007}. For M33, the stellar 
halo is significantly more metal poor ([Fe/H]$\simeq-1.5$) than would be inferred 
from the metallicity of the disk stars ([Fe/H]$\simeq-0.9$, \citealt{ref:McConnachie_2006}). 
These results provide strong circumstantial evidence supporting our conclusion 
that the SCSNe explosions far from the center of their hosts imply that their 
progenitors were located in metal-poor environments, likely linked to 
low-metallicity star formation not associated with central regions of their 
hosts. 

Beyond the Local Group, discovery of extended UV disks around nearby galaxies 
M83~\citep{ref:Thilker_2005} and NGC~4625~\citep{ref:Gil_2005} have shown that 
while there is evidence for ongoing star formation in the very outer edges of 
these galaxies, both the star formation rate and total stellar mass in these 
regions are very small. Dwarf galaxies and extended disks 
contain only a few percent of the total stellar mass or star formation 
\citep[e.g.,][]{ref:Benson_2007}, and in the local universe about half of the 
stellar mass is contained in elliptical galaxies or the bulges of large spirals 
\citep[e.g.,][]{ref:Tasca_2011}. Yet, all four of the confirmed SCSNe exploded 
in these environments. Even including the additional candidate SCSNe considered by 
\cite{ref:Taubenberger_2011}, for which the evidence of their SCSN nature 
is less conclusive, none has exploded in an early type galaxy. In other words, 
what is most striking about the SCSNe explosion sites is not where they are 
located, but rather where they are not: they avoid both spiral galaxy disks and 
elliptical galaxies, even though these are the regions that contain almost all 
the stellar mass and star formation. This is highly unlikely to be a selection 
effect, given that the SCSNe are some of the brightest SNe~Ia ever detected, and 
thus are easier to discover in SN surveys than ordinary SNe~Ia.

The four confirmed SCSNe events are the high luminosity tail of the observed 
SNe~Ia luminosity distribution, and thus by inference, total WD mass. This raises 
the possibility that other slightly less luminous SNe~Ia may also have 
resulted from double-degenerate mergers resulting in progenitor masses exceeding 
the Chandrasekhar limit by a narrower margin. As we 
discussed in \cite{ref:Khan_2011}, these bright SNe~Ia also tend to lie far from 
the centers of their hosts or in dwarf galaxies, suggesting more generally that 
their higher luminosities are likely related to the lower metallicities of their environments,
although whether this is driven by age or metallicity is hotly debated 
\citep[e.g.,][]{ref:Hamuy_2000,ref:Gallagher_2005,ref:Gallagher_2008,ref:Prieto_2008,ref:Howell_2009}. 
For example, Figure~21 of \cite{ref:Hicken_2009} shows that roughly a third of 
all SNe~Ia have a projected distance of $\gtrsim10$~kpc from the centers of hosts. 
Some specific examples include SN~2007bk, where the over-luminous SN was found 
$\sim9$~kpc away from its metal-poor ($Z \simeq 0.25 Z_\odot$) dwarf host~\citep{ref:Prieto_2008}, 
and the 1991T-like SN~2009nr which is located 4.3 disk scale lengths away from 
its host~\citep{ref:Khan_2011}.

It is possible that in metal-poor systems the WD Initial-Final Mass Relation is highly 
unconventional and the low metallicity systems produce an unexpected 
overabundance of SCSNe~\citep[e.g.,][]{ref:Umeda_1999a}. Metallicity may also significantly contribute to the SN~Ia 
delay time distribution~\citep[e.g.,][]{ref:Meng_2011c}.
Alternatively, the mapping of total WD mass to the 
expected $^{\rm 56}$Ni yield in the DD scenario~\citep[e.g.,][]{ref:Howell_2009} 
may be especially problematic in the very metal-poor tail of WD populations, 
and comparable WD masses may lead to a wide range of luminosities~\citep[e.g.,][]{ref:Umeda_1999b,ref:Chen_2009,ref:Pakmor_2010}. 
In the SD scenario, differential rotation of the WD has been proposed as a means of 
exceeding the Chandrasekhar-mass limit~\citep{ref:Piro_2008} and low-metallicity 
may be required for SD progenitors to highly surpass this threshold~\citep{ref:Hachisu_2011}.
The small sample size of four confirmed SCSNe limits our ability to verify any of these 
hypotheses, but the discovery of more SCSNe and re-analysis of previously discovered very 
luminous SNe~Ia should lead to a deeper understanding of the underlying 
systematics affecting SNe~Ia luminosities, rates, and the nature of their 
progenitors.

\acknowledgments
We thank A.~Gould for his ardent encouragement, C.~S.~Kochanek for useful suggestions, 
and the anonymous referee for helpful comments. 
We are grateful to the staffs of the MDM Observatory for their excellent support. 
This research has made use of NED, which is 
operated by the JPL and Caltech, under contract with NASA and the HEASARC 
Online Service, provided by NASA's GSFC.
Funding for the SDSS and SDSS-II has been provided by the Alfred P. Sloan Foundation, 
the Participating Institutions, the National Science Foundation, the U.S. Department 
of Energy, the National Aeronautics and Space Administration, the Japanese Monbukagakusho, 
the Max Planck Society, and the Higher Education Funding Council for England. 
This research has made use of 
photographic data of the National Geographic Society -- Palomar Observatory Sky 
Survey (NGS-POSS) obtained using the Oschin Telescope on Palomar Mountain. The 
NGS-POSS was funded by a grant from the National Geographic Society to the 
California Institute of Technology. The Digitized Sky Survey was produced at 
the Space Telescope Science Institute under US Government grant NAG W-2166. RK 
and KZS are supported in part by NSF grant AST-0707982. KZS is 
supported in part by NSF grant AST-0908816. 
RS is supported by the David G. Price Fellowship in Astronomical Instrumentation.
JLP acknowledges support from NASA through Hubble Fellowship grant HF-51261.01-A 
awarded by the STScI, which is operated by AURA, Inc. for NASA, under contract NAS 5-26555.

\end{document}